\documentclass{mn2e}

\usepackage{epsfig,amsmath,amssymb,amsfonts,mathrsfs}
\RequirePackage{amssymb}
\RequirePackage{amsfonts}

\begin{document}

\newcounter{count}

\title{Non-linear radial oscillations of neutron stars: Mode-coupling results}

\author[U.~Sperhake, P.~Papadopoulos, N.~Andersson]{U.~Sperhake$^{(1)}$, 
P.~Papadopoulos$^{(2)}$, N.~Andersson$^{(1)}$\\
        (1) Department of Mathematics, University of Southampton, Southampton 
SO17 1BJ, United Kingdom \\
        (2) School of Computer Science and Maths, University of Portsmouth, 
        Portsmouth PO1 2EG, United Kingdom}

\date{Accepted.
      Received;
      in original form October 2001}

\maketitle

\begin{abstract}
The non-linear behaviour of oscillation modes in compact stars is a
topic of considerable current interest. Accurate numerical studies of
such phenomena are likely to require powerful new approaches to both
fluid and spacetime computations. We propose that a key ingredient of
such methods will be the non-linear evolution of deviations from the
background stationary equilibrium star. We investigate the feasibility
of this approach by applying it to non-linear radial oscillations of a
neutron star, and explore numerically various non-linear features of
this problem, for a large range of amplitudes. Quadratic and higher
order mode coupling and non-linear transfer of energy is demonstrated
and analysed in detail.
\end{abstract}

\section{Introduction}
Since the first realisation that the Cepheids are stars that pulsate
radially, stellar oscillations have provided a fertile ground for
astrophysicists. Observations obtained with much improved sensitivity
have provided a wealth of relevant data in recent years. This provides
an important theoretical challenge given that the gathered information
can be matched against detailed stellar models. With the likely advent
of gravitational-wave astronomy in the next few years, relativists are
considering whether a similar program may be feasible for compact
stars. This is an exciting idea since the detection of gravitational
waves from a pulsating star may shed light on the nature of the
equation of state at supra-nuclear densities.  Although stellar
oscillation theory has been an active field of research for many
decades (in particular in the context of Newtonian gravity) and there
are several monographs covering the main results, many crucial
questions remain open. The main uncertainties concern the behaviour in
the non-linear regime, e.g., the coupling between different modes, the
formation of shocks etcetera. The purpose of the present paper is to
demonstrate the accuracy of a new approach to the study of non-linear
stellar oscillations.  We apply the new method, in which the main
focus is on non-linear deviations from the background stationary
equilibrium star, to radial oscillations of neutron stars.  This is a
problem which, given its significance for the stability of the star,
has received a lot of attention in the past.  Although most results
have been obtained in the linear regime, and concern the nature of the
various eigenmodes \cite{chan77,gl83,vc92,kr01}, there have also been
attempts to study non-linear features in a perturbative way (including
quadratic and cubic coupling terms)
\cite{dzie82,pb82,perd83,went87,kg89,vh96}. [It is also relevant to
mention the recent application of this approach to non-linear effects
on inertial modes \cite{saftw01} as well as the fully nonlinear simulations by Font, Stergioulas and Kokkotas \shortcite{font}.] Our new approach provides a
powerful complement to these studies.  We investigate the main
features that appear in the weak to mildly non-linear regimes. This
leads to some interesting new results regarding non-linear
mode-coupling and sheds light on two of the main questions in this
area: i) What is the amplitude at which large amplitude modes
saturate?, and ii) How reliable are expansion methods beyond quadratic
order in the amplitude?  We believe our paper provides the first
detailed study of these effects in full general relativity.

\section{Non-linear perturbations}
We model the neutron star as a single component perfect fluid at zero
temperature which obeys a polytropic equations of state $ P =
K\rho^{\gamma}$, where $K$ and $\gamma$ are constants. For such a
fluid the energy momentum tensor is given by $ T^{\mu \nu} = (\rho +
P) u^{\mu} u^{\nu} + P g^{\mu \nu}$, where $u^{\mu}$ is the four
velocity, normalised as $u^{\mu}u_{\mu}=-1$.  We restrict our
consideration to spherical stars undergoing radial motions, in which
case the four velocity is given by
$u^{\mu}=[v(t,r),w(t,r),0,0]$. Furthermore, in radial gauge and polar
slicing the spherically symmetric line element is,
\begin{equation}
  ds^2 = -\lambda^2 dt^2 + \mu dr^2 + r^2(d\theta^2 + \sin{\theta})
      d\phi^2 \ ,
\end{equation}
i.e. it depends only on two
functions $\lambda(t,r)$ and $\mu(t,r)$. With those assumptions the
Einstein field equations $G_{\mu \nu} = 8\pi T_{\mu \nu}$ and the
equations of hydrodynamics $\nabla_{\mu} T^{\mu}{}_{\nu}=0$ lead to
four independent equations
\begin{eqnarray}
  &\frac{\lambda_{,r}}{\lambda} = \frac{\mu^2-1}{2r} + 4\pi r
      \mu^2 \left[ P + (\rho + P)
      \,\mu^2w^2 \right], \label{NONP_LAMBDAR} \\[10pt]
  &\frac{\mu_{,r}}{\mu} = -\frac{\mu^2-1}{2r} + 4\pi r
      \mu^2 \left[ \rho + (\rho + P)
      \mu^2w^2 \right], \label{NONP_MUR} \\[10pt]
  &\rho_{,t} + \alpha_{11} \rho_{,r} + \alpha_{12} w_{,r} = b_1,
  \label{NONP_RHOT}  \\[10pt]
  &w_{,t}          + \alpha_{21} \rho_{,r} + \alpha_{11} w_{,r} = b_2.
  \label{NONP_WT}
\end{eqnarray}
Here $\alpha_{11}$, $\alpha_{12}$, $\alpha_{21}$, $b_1$ and $b_2$ (see
Sperhake \shortcite{sper01} for their exact form) are functions of the
fundamental variables $\lambda$, $\mu$, $\rho$ and $w$, but not of
their derivatives, so that Eqs.\,(\ref{NONP_RHOT}), (\ref{NONP_WT})
form a quasi-linear system.  The solution of
Eqs.\,(\ref{NONP_LAMBDAR})-(\ref{NONP_WT}) requires suitable boundary
conditions. At the centre of the star we require that $\mu=1$ in order
to avoid a conical singularity. The velocity $w$ vanishes at the
origin due to spherical symmetry. At the stellar surface we fix the
lapse function by matching the line element to an exterior
Schwarzschild metric which implies that $\lambda\mu=1$.  We use
typical parameters for the neutron star model: $K=150\,\,{\rm km}^2$
and $\gamma=2$. We set the central density of the equilibrium
configuration of the neutron star to $\rho_{\rm c}=0.001224\,\,{\rm
km}^{-2}$, which leads to a compactness (mass/radius ratio)
$M/R=0.19$.

In our analysis of non-linear mode coupling we approach radial
oscillations of neutron stars satisfying equations
(\ref{NONP_LAMBDAR})-(\ref{NONP_WT}) in the
following manner. We decompose the time dependent quantities
$\lambda$, $\mu$, $\rho$ and $P$ into static
background contributions and time dependent perturbations according to
\begin{equation}
	f(t,r)=f(r) + \delta f(t,r).
\end{equation}
The background quantities are determined by the
Tolman-Oppenheimer-Volkoff equations, the static analogues of
Eqs.\,(\ref{NONP_LAMBDAR})-(\ref{NONP_WT}). We use the background
equations to eliminate all zero order terms from the resulting
perturbative dynamic equations.  The resultant equations
are equivalent to the original system
(\ref{NONP_LAMBDAR})-(\ref{NONP_WT}). In particular, the two evolution
equations form a quasilinear system for $\delta \rho$ and $w$.  
By eliminating terms of order zero in the perturbations we
obtain  numerical accuracy 
that is determined by the amplitude of the perturbation rather than
the static background. This is the key advantage of this new approach.
All preliminary tests have verified that this non-linear perturbation
scheme provides much enhanced accuracy over a large range of
amplitudes. Full details of the new method as well as the 
numerical code and its calibration are
provided by Sperhake~\shortcite{sper01}.

Normally the surface of the star is defined by the vanishing of the
pressure, which in the polytropic case is equivalent to
$\rho=0$. However, if one is using an Eulerian framework and the
surface of the star is allowed to move, the outer grid boundary does
not coincide with the surface of the star and this condition cannot be
applied easily. As has been discussed in detail by
Sperhake~\shortcite{sper01} this leads to severe numerical
difficulties, and could trigger artificial shock formation in the
surface region.  In the present study we want to focus on the
non-linear coupling between various oscillation modes. In order to
isolate this effect (and avoid any artificial effects due to the
surface of the star) we use a fixed (rather than free) boundary
condition, i.e. we require $w=0$ at the surface. Furthermore we do not
evolve the low density layers of the neutron star, in order to avoid
negative total energy densities. The resulting neutron star model
contains about 90 $\%$ of the mass of the original model.

\begin{figure}
  \centering
  \epsfig{file=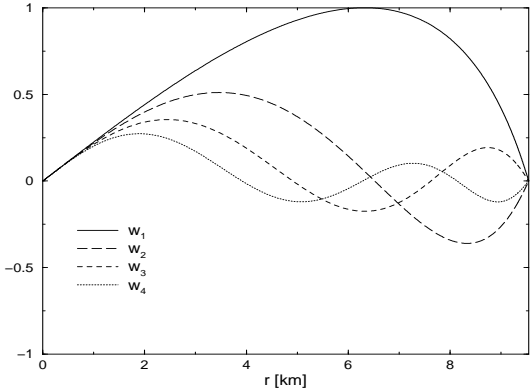, height=200pt, width=150pt, angle=-90}
  \caption{The profiles of the lowest four eigenmodes in the
           velocity $w$ for the neutron star model with fixed surface.
       }
  \label{MOD_EIG}
\end{figure}

The eigenmodes of a dynamic spherically symmetric neutron star are
described by the linearised version of the dynamic equations
(\ref{NONP_LAMBDAR})-(\ref{NONP_WT}). It is a well known result that
the linearised equations lead to a self adjoined eigenvalue problem in
terms of the rescaled displacement vector $\zeta$, which is related to
our variables by $ \zeta_{,t} = r^2 w$, cf. chapter~26 in Misner,
Thorne and Wheeler \shortcite{mtw}.  The solutions of the eigenvalue
problem form a complete orthonormal system and hence we can
expand the time dependent $\zeta(t,r)$ resulting from fully
non-linear evolution in a series of the linear eigenmodes
\begin{align}
  \zeta(t,r) &= \sum_i{A_i(t)\zeta_i(r)},
\label{modesum}\end{align}
where the coefficients $A_i$ are given in terms of the inner product
\begin{equation}
  A_i(t) = \int_0^R(P+\rho)
{ \mu^3 \lambda \over r^2} \zeta(t,r)\zeta_i(r) dr  \,.
\label{ortho}\end{equation}
In Fig.\,\ref{MOD_EIG} we show the four lowest eigenmodes in the
velocity field for our truncated stellar model. The expansion
(\ref{modesum}) provides a useful diagnostic which allows us to assess
the level of non-linear coupling throughout a numerical evolution.  We
measure the presence of mode $i$ at any given time during the
evolution by calculating the coefficient $A_i(t)$ via numerical
integration.

\section{Results}
In order to study non-linear mode-coupling we evolve the nonlinear
perturbation equations from initial data corresponding to the velocity
field $w_j$ of a single linear eigenmode (of order $j$) with amplitude
$K_j = \hbox{max} |\lambda \zeta_j(r)/r^2|$, which represents the
maximum radial displacement of a fluid element inside the star.  The
initial density perturbation $\delta \rho$ is set to zero, while the
initial values for the metric variables follow from the constraints
(\ref{NONP_LAMBDAR})-(\ref{NONP_MUR}).

\subsection{Exciting the fundamental mode}
\begin{figure}
  \centering
  \epsfig{file=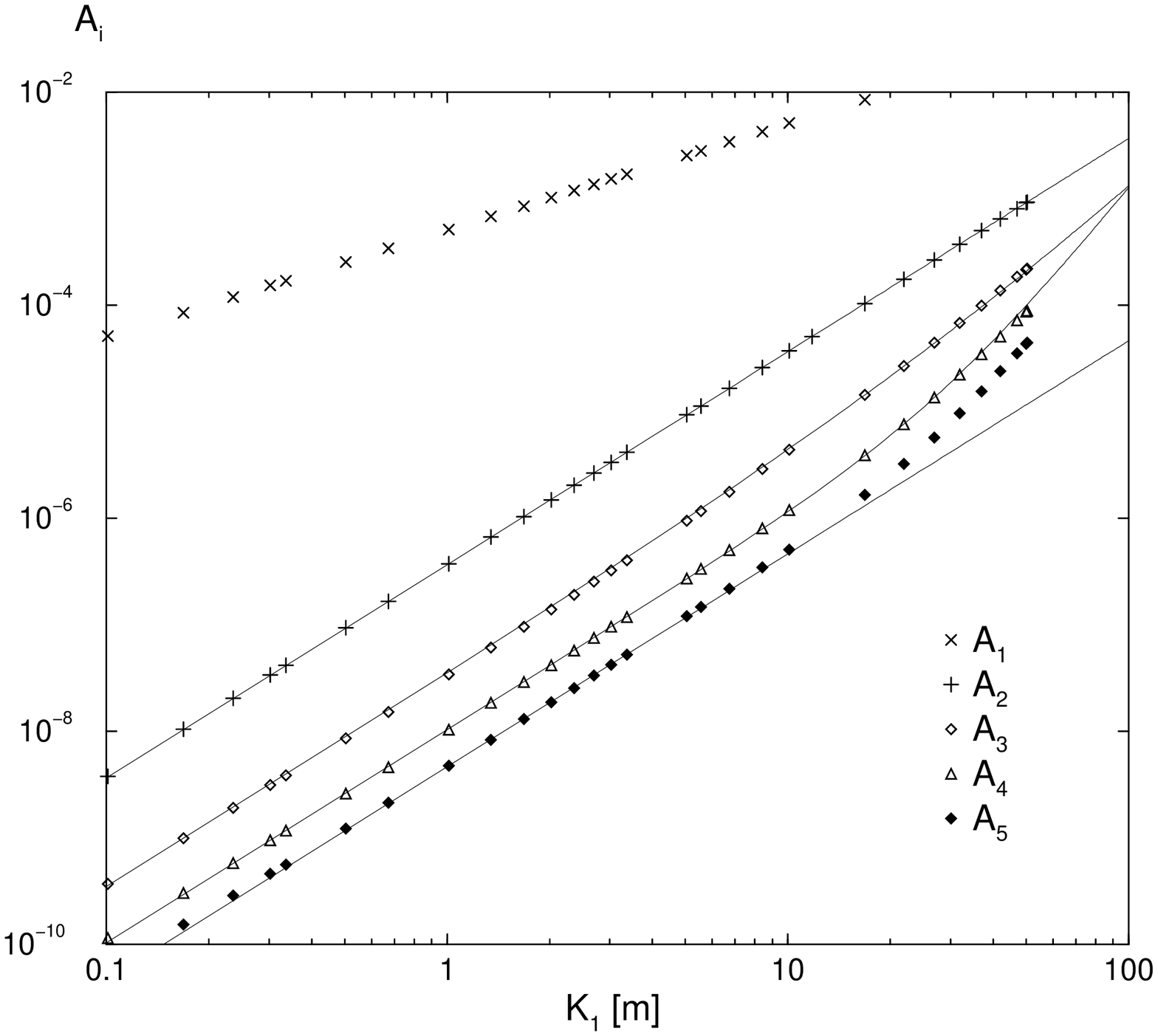, height=225pt, width=175pt, angle=-90}
  \epsfig{file=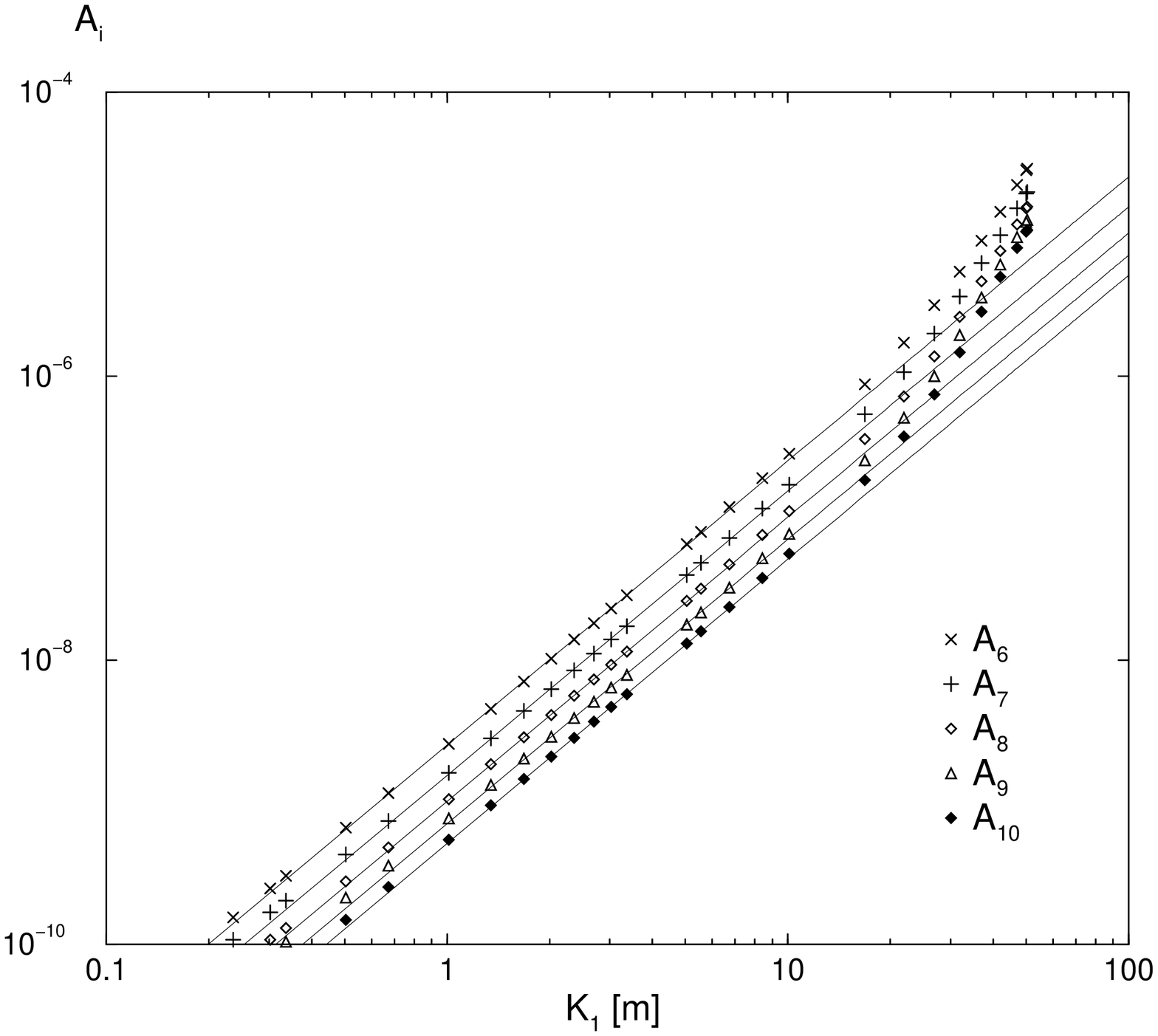, height=225pt, width=175pt, angle=-90}

  \caption{The coupling of the fundamental mode to higher order modes
           is illustrated by plotting the coefficients $A_i$ as
           functions of the initial amplitude $K_1$. For low
           amplitudes the results are well fitted by quadratic power
           laws  while higher powers
           are required in the mildly non-linear regime 
(as indicated by the solid lines).}
           \label{SQUARE01}
\end{figure}

We first consider the case when the initial data correspond to the
fundamental radial eigenmode.  Having evolved this data, we measure
the maximal coefficients $A_i= \max|A_i(t)|$ obtained over an
integration time corresponding to many times the dynamical timescale.
In Fig.\,\ref{SQUARE01} we show the maximal coefficients obtained for
the lowest 10 eigenmodes for excitation amplitudes ranging between
$1\,\,{\rm cm}$ and $50\,\,{\rm m}$. We observe weak mode-coupling
throughout most of this domain.  The fundamental mode itself ($A_1$),
is seen to grow more or less linearly with the initial amplitude
$K_1$, which indicates the absence of significant self interaction.
Meanwhile, for higher order modes we can clearly identify two
different regimes: For amplitudes below $10\,\,{\rm m}$, all
coefficients $A_2,\ldots,A_{10}$ grow quadratically with the
excitation amplitude $K_1$. At larger amplitudes all eigenmode
coefficients except for $A_2$ show a transition to power laws with
larger index. We have illustrated this behaviour in
Fig.\,\ref{SQUARE01} by modelling the coefficients $A_2$, $A_3$ and
$A_4$ as power series expansions in $K_1$ according to
\begin{equation}
A_i=c_i K_1^{\,2}+d_i K_1^{\,i},
\end{equation}
(the Einstein summation convention is not use here or in similar
expressions below) where $c_i=\{ 3.6\cdot10^{-7},\, 3.4\cdot
10^{-8},\, 1.0\cdot 10^{-8} \}$ and $d_i=\{ 0,\, 9.7\cdot 10^{-10},\,
1.2\cdot 10^{-11} \}$ for $i=2,3,4$.  The higher order power laws have
been obtained by least square fits to the coefficients $A_i$ after
subtracting the quadratic contributions $c_i K_1^{\,2}$.  For the
modes $i=5-10$ the contribution of the higher order power law is
rather weak which makes it difficult to obtain accurate measurements
of the corresponding exponents. The steepening of the curves
is, however, still obvious in the figure.  It is also clear, since the curves
for the eigenmode coefficients $A_i$ do not intersect in
Fig.\,\ref{SQUARE01}, that the coupling strength decreases with the
order of the eigenmodes over the whole range of amplitudes.

\subsection{Exciting higher order modes}

\begin{figure}
  \centering \epsfig{file=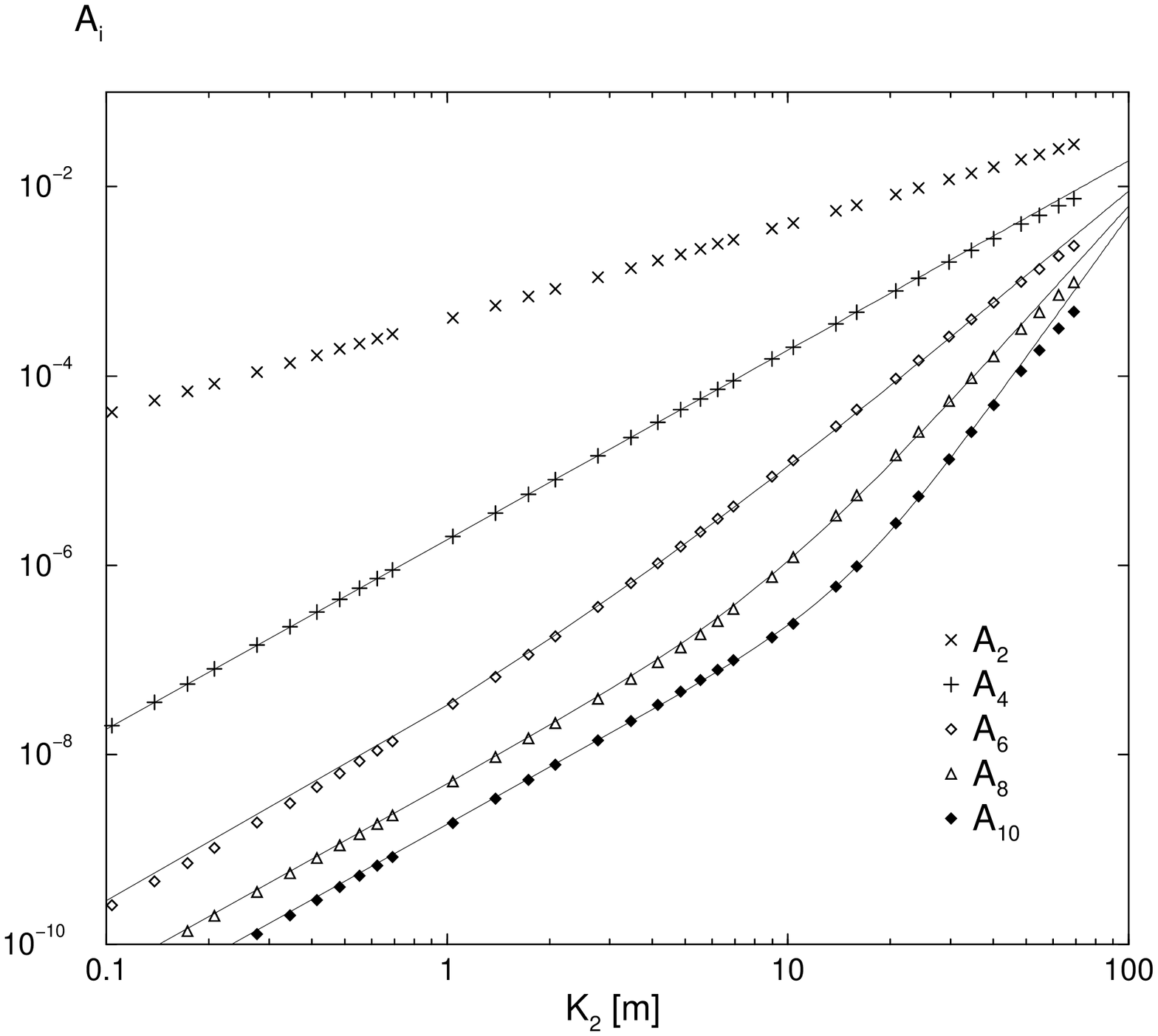, height=225pt, width=175pt,
  angle=-90} \epsfig{file=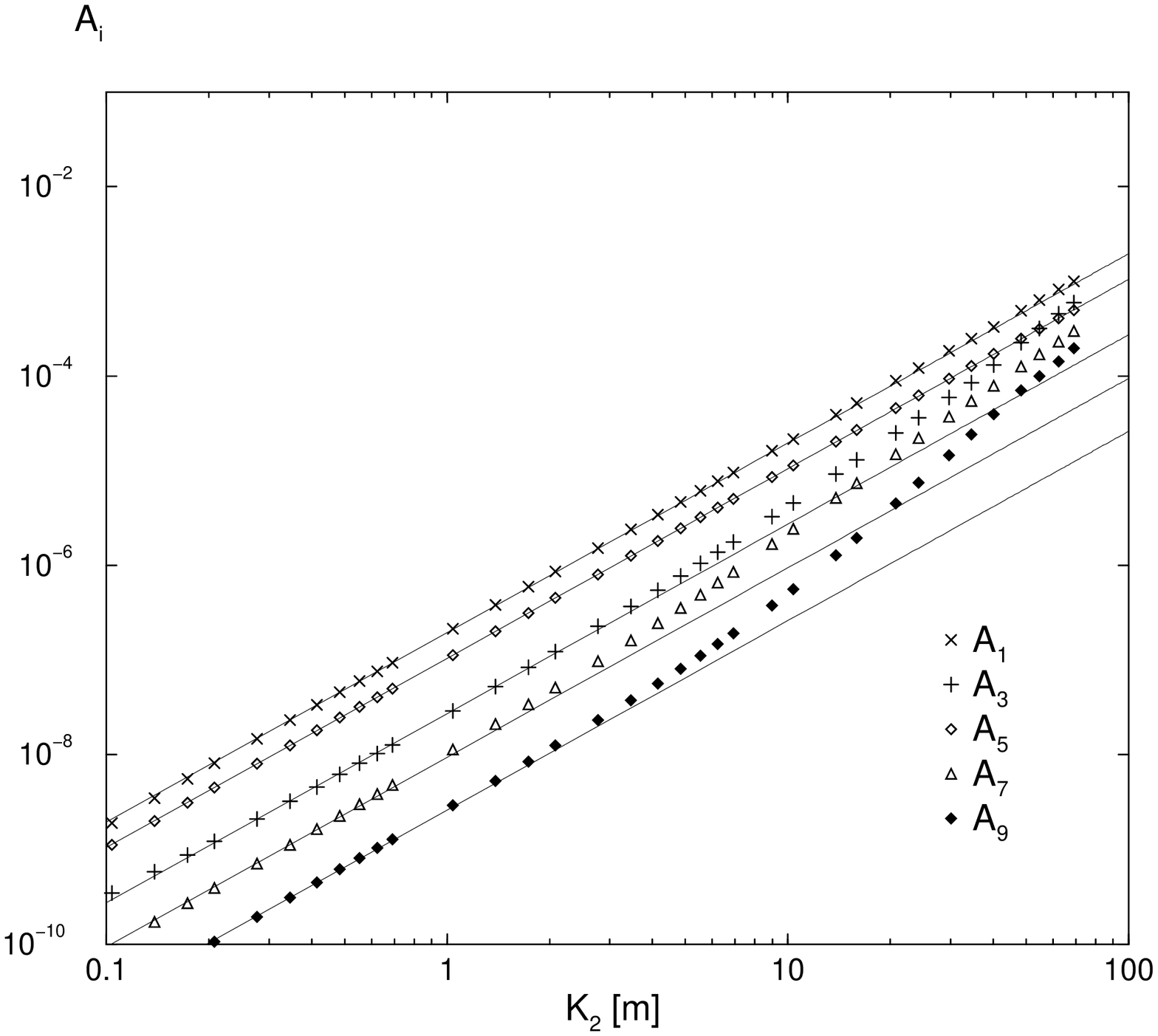, height=225pt, width=175pt,
  angle=-90} \caption{The eigenmode coefficients obtained for initial
  data in the form of the second eigenmode. We have fitted the data in
  the intermediate amplitude range between $K_1=1\,\,{\rm m}$ and
  $10\,\,{\rm m}$ by quadratic power laws.  The top panel shows the
  resulting fits to $c_{i}\cdot K_2^{2}+d_{i}\cdot K_2^{\,i/2}$ }
  \label{SQUARE02}
\end{figure}

Next we set initial data in the form of second and higher eigenmodes.
In this case we still observe the two regimes mentioned above.  In the
weakly non-linear regime the eigenmode coefficients are well
approximated by quadratic power laws. The main difference from the
previous case is that the second mode (of order $i$) shows a
preference to couple to modes of order $2i$. This is particularly clear in
Fig.\,\ref{SQUARE02} where the coefficients $A_6$, $A_8$ and $A_{10}$
show a much steeper increase for large amplitudes. For these modes we
also obtain excellent fits with combinations of power laws according
to
\begin{equation}
A_{i}=c_{i} K_2^{2}+d_{i}K_2^{\,i/2}
\end{equation}
where $c_i =
\{2.9\cdot10^{-6},\,2.5\cdot 10^{-8},\,4.9\cdot 10^{-9}, \,1.8\cdot
10^{-9} \}$ and $d_i = \{0,\,8.7\cdot10^{-9},\,6.2\cdot 10^{-11},
\,4.9\cdot 10^{-13} \}$ for $i=4, 6, 8, 10$. These fits are shown in
Fig.\,\ref{SQUARE02}.

These results are generally confirmed if the initial data is given in
the form of the third velocity mode. The only difference is that the
preferred modes in the moderately non-linear regime are now those of
order $3i$.

\section{Discussion}
In this paper we have applied a new non-linear approach to the study
of stellar pulsation, and studied mode-coupling due to non-linear
effects by evolving initial data corresponding to a single linear
eigenmode with varying amplitude. Concerning the transfer of energy to
other modes we have found two distinct regimes, a weakly non-linear
regime where the excitation of modes grows quadratically with the
initial amplitude and a moderately non-linear regime, which can be
reasonably well described by power laws of higher order.

The results for the weakly non-linear regime agree qualitatively with
Newtonian perturbative studies.  In the analytic study of non-linear
mode coupling one normally views the eigenmode coefficients as
harmonic oscillators and the non-linear interaction between eigenmodes
is represented in the form of driving terms which are quadratic or of
higher order in the amplitudes (see for example van
Hoolst~\shortcite{vh96})
\begin{align}
  \frac{d^2 A_i}{dt^2} + \omega_i^2 A_i &= c_i^{jk} A_jA_k + d_i^{jkl}
      A_j A_k A_l + \ldots,
      \label{FORCEDOSCILLATOR}
\end{align}
where the $c_i^{jk}$, $d_i^{jkl},\ldots$ are the quadratic, cubic and
higher order coupling coefficients and summation over $j,k,l$ is
assumed. In our analysis the initial data consists of one isolated
eigenmode $j$, so that the right hand side can be approximated by $c_i
A_j^2 + d_i A_j^3 + \ldots$ In analytic studies this series expansion
is normally truncated at second or third order.  The omission of
higher order terms is justified in the weakly non-linear regime, where
our fully non-linear simulations confirm that quadratic terms in the
initial amplitude dominate the coupling between eigenmodes.  This is
no longer true, however, in the moderately non-linear regime, where
higher order terms are more important. Our results allow us to define
the transition to this regime, manifested by the breaks in the curves
in Figures~\ref{SQUARE01} and \ref{SQUARE02}. As is clear from the
figures, the moderately non-linear regime corresponds to initial mode
amplitudes above 10~m or so. We note that the corresponding Mach
number is of the order of 0.01. This agrees well with investigations
of Newtonian stars (Kumar \& Goldreich 1989) which assume a Mach
number of 0.1 as the limit of applicability of semi-analytic mode
coupling methods.

Furthermore, our results indicate that the non-linear couplings would
be poorly captured by polynomial expansions in the mildly non-linear
regime.  We observe significant excitation of higher order modes (eighth,
tenth etc) which can only emerge from very high-order couplings.  To
quantify the associated coupling coefficients in a perturbative
calculation would be very difficult.

We have also observed that, given an initial mode $j$, the coupling to
modes $nj$ is particularly efficient in the moderately non-linear
regime. This is naturally interpreted as a resonance effect. In
analogy with the simple problem of a single forced oscillator we can
assume that resonance occurs for any mode whose frequency is an
integer multiple of the driving frequency $\omega_j$ in the general
non-linear case, i.e.  we can schematically write the eigenmode
coefficients in the form
\begin{align}
  A_i(t) &\sim \sum_n \frac{F_n}{\omega_i^2 - (n\omega_j)^2}, \label{RESONANCE}
\end{align}
where the $F_n$ will depend on the frequencies (cf. Eqs.\,(18), (19)
of van Hoolst~\shortcite{vh96}. In our case
the external force is provided by the
non-linear coupling to the initial mode $j$, as indicated in
(\ref{RESONANCE}). We therefore obtain resonance if $\omega_i = n
\omega_j$. For our simple neutron star model, the eigenfrequencies of
radial neutron star oscillations are almost equally spaced and we can
use $\omega_j\approx (j \omega_i)/i$ for $i,j\ge 2$ as a reasonable
approximation. The condition for resonance then becomes $i = n j$
which is exactly what we have observed.

As one of the main results of this paper, we emphasise the new
perturbative approach that enabled us to obtain highly accurate, fully
non-linear, evolutions over a large range of amplitudes.  This
approach was discussed in detail by Sperhake \shortcite{sper01}, and
we plan to present further results regarding mode-coupling and
non-linear shock formation in future papers. In principle, this
technique can be applied to any physical problem that involves a
non-trivial stationary limit and we expect it to prove a valuable tool
in many non-linear problems. We note that the accuracy improvements
are independent of the numerical discretization used (here, second
order, centred finite differences). In combination with methods
suitable for smooth oscillatory solutions~\cite{gour91,ghg95}, we
would expect a dramatic expansion of the applicability of non-linear
simulations to relativistic stellar pulsations.

A problem for which our new approach may prove useful concerns
unstable modes of rotating neutron stars (eg. the r-modes, see
Andersson \& Kokkotas \shortcite{ak01} for a review). One of the most
important questions raised in connection with the r-modes concerns
the amplitude at which an unstable mode saturates.  Recently, direct
three dimensional numerical simulations have been brought to bear on
this problem \cite{sf01,ltv01}. The picture that emerges from these
studies is, however, not conclusive. Both studies suggest that an
unstable r-mode saturates at an extremely large amplitude 
(corresponding to waves of a height of several hundred meters in a star
spinning near the breakup limit) due to shocks forming in the surface
region.  That such ``wave breaking'' would occur once a mode reaches a
large amplitude is likely, but simulations must isolate the true physical
behaviour near the surface of a star, from numerical artifacts
associated with the rapid decrease in the density (for a detailed
discussion see Sperhake \shortcite{sper01}).  In fact, it is
interesting to contrast these results with those of the present work
that suggest that nonlinear effects are highly relevant already at
wave amplitudes of order 10~m. We believe that a suitable
generalisation of the method used in this paper, that provides
unprecedented accuracy for a large range of wave amplitudes, could
prove extremely useful for the study of unstable non-axisymmetric
modes and plan to address such problems in the near future.

\subsection*{Acknowledgements}
We thank Kostas Kokkotas for helpful comments. 
This work has been supported in part by the EU Programme
'Improving the Human Research Potential and the Socio-Economic
Knowledge Base', (Research Training Network Contract
HPRN-CT-2000-00137). P.P. acknowledges support from the Nuffield
Foundation (award NAL/00405/G). N.A. acknowledges support from the Leverhulme
Trust in the form of a prize fellowship.


\end{document}